\documentclass[11pt]{article}
\usepackage[dvips]{pstcol} 
\usepackage[dvips]{graphicx}
\usepackage{psfrag}
\def\today{\ifcase \month\or
  January\or February\or March\or April\or May\or June\or
    July\or August\or September\or October\or November\or December\fi
      \space\number\day,\space
        \number\year }
\newcommand{\be}{\begin{equation}}
\newcommand{\en}{\end{equation}}
\newcommand{\bea}{\begin{eqnarray}}
\newcommand{\ena}{\end{eqnarray}}
\newcommand{\lbl}[1]{\label{eq:#1}}
\newcommand{\rf}[1]{(\ref{eq:#1})}
\newcommand{\lapprox}{%
\mathrel{%
\setbox0=\hbox{$<$}\raise0.6ex\copy0\kern-\wd0\lower0.65ex\hbox{$\sim$}}}
\def\e{ {\rm e}}
\def\tr{ {\rm tr\,}}
\newcommand{\trace}[1]{\langle #1 \rangle}
\begin{document}
\begin{flushright}
%\today 
\end{flushright}
\begin{center}
{\Large\bf Chiral perturbation theory:\\
a basic introduction\footnote{
Lectures given at the FANTOM study week, Emmen, May 24-28, 2004. 
Work supported in part by IFCPAR contract 2504-1 and by the European
Union TMR network contract HPRN-CT-2002-00311 (EURIDICE). }}
\vskip 1cm

B. Moussallam\\
{\sl  Groupe de Physique Th\'eorique, IPN }\\
{\sl Universit\'e Paris-Sud}
{\sl F-91406 Orsay C\'edex, France}
\vskip 0.6cm

{\Large\bf Abstract}\\
\vskip 0.3cm

\begin{minipage}{13cm}
Chiral perturbation theory is a very general expansion method which can be
applied to any dynamical system which has continuous global symmetries
and in which the ground state breaks some of these spontaneously. In these
lectures we explain at a basic level and in detail how such symmetries 
are identified in the case of the QCD Lagrangian and describe the steps
which are involved in practice in the construction of a low-energy
effective theory for QCD.
\end{minipage}
\end{center}

\vskip 1cm
\section{Introduction}
QCD is a theory which gives rise to extremely diverse phenomena depending
on the energy range which is involved. At low energies ($E\lapprox 1$ GeV)
the physics is non perturbative and it is in many aspects 
strongly influenced by chiral symmetry. 
Chiral symmetry is a global symmetry of the QCD Lagrangian (as we will
discuss in detail) in the case where $N^0_F\ge 2$ quarks are massless and 
it happens 
that the vacuum of this theory breaks the symmetry spontaneously, a fact which
is linked to  confinement.
Physically, this property is relevant because some of the quarks happen
to be very light, with a mass much smaller than 1 GeV,  
such that they do probe, in a sense, the chiral vacuum.
The method of chiral perturbation theory (ChPT) 
exploits this property of QCD in a general and systematic way. 
These  lectures are an introduction to this subject at a  basic level
which could help the reader in tackling, later on, the classic papers on this 
subject\cite{weinberg}\cite{gl84}\cite{gl85}. For a rather complete
introduction one may consult the recent review\cite{scherer}. A detailed 
description of applications can be found in the book\cite{dghbook} and, for
the quantum field theoretical aspects, of course, one should consult 
Weinberg's book\cite{weinbook}.

\section{Symmetries of chiral QCD}
The field content of QCD consists of  a set of 8 massless gluon fields 
$G^a_\mu$ associated with the $SU_c(3)$ colour gauge group and a set of 
six quarks $q_f$. The mass pattern of the quarks is remarkable and is 
displayed below

\begin{center}
\begin{tabular}{lll}
$f$    &   $q_f$   &  $m_f$ ( MeV) \\
1    &    u    &  $\sim$ 5   \\
2    &    d    &  $\sim$ 7   \\
3    &    s    &  $\sim$ 100   \\
4    &    c    &  $\sim$ 1200   \\
5    &    b    &  $\sim$ 4200   \\
6    &    t    &  $\sim$ 174000  
\end{tabular}  
\end{center}
The masses are ``running'' parameters i.e they depend on a scale $\mu$
(for the light quarks the numbers above correspond to $\mu=2$ GeV). 
At present, the determination of the masses is 
{\sl very} approximate (see e.g.\cite{pdg}).
One of the goals of the chiral expansion is to extract informations on the
values of lightest three quarks from experiment. The numbers above 
suggest to divide the quarks into a group of ``light'' quarks, $f\le 3$
and a group of heavy quarks for $f\ge 4$. The chiral expansion is an effective 
Lagrangian  technique which generates an expansion in powers of the light
quark  masses, $m_f$ divided by a scale $\Lambda$ which is expected to
be of the order of 1 GeV. This expansion is completely non perturbative
with respect to the QCD running coupling constant $g_s$ and will necessitates
that one makes a change from the quark and gluon fields to a new set of 
variables. 

\subsection{Identification of the symmetries}
As a first step, let us write the QCD Lagrangian in the following form
\be
{\cal L}_{QCD}= {\cal L}^0_{QCD} + {\cal L}^{sources}_{QCD}
\en
with
\bea
&&{\cal L}^0_{QCD}= -{1\over4} G^a_{\mu\nu} G^{a\,\mu\nu} +
\sum_{f={N^0_F+1}}^6
\bar q_f (i\gamma ^\mu D_\mu- m_f ) q_f\nonumber\\
&&\phantom{{\cal L}^0_{QCD}=  }
+ \sum_{f=1}^{N^0_F} \bar q_f\gamma ^\mu (iD_\mu ) q_f\ .
\ena
Here, $D_\mu$ is the covariant derivative
\be
iD_\mu =i\partial_\mu +g_s G^a_\mu {T^a}\ .
\en
with $T^a$ being the eight generators of the $SU_c(3)$ colour group
(i.e. $T^a=\lambda^a/2$, $\lambda^a$ are the Gell-Mann matrices). 
The quark fields must be considered as 3-vectors in colour space. 
In this formula, 
the quark flavours with $f\ge {N^0_F+1}$ are the ones which
we decide to treat as heavy. 
We note that ${\cal L}^0_{QCD}$ does not contain any mass term for 
the quarks  with flavour index $f\le {N^0_F}$, which we will call
the light quarks. There is some freedom in the choice  
of ${N^0_F}$, one can take ${N^0_F}=2$ or ${N^0_F}=3$. For definiteness,
we will assume ${N^0_F}=3$ in the following.
The mass terms of the light quarks are included in the second
piece ${\cal L}^{sources}_{QCD}$  which we will discuss in detail later. 
This piece will also account for the couplings of the quarks to the other
fields in the standard model.
The piece ${\cal L}^0_{QCD}$ is to be treated exactly: the dynamics will
be reflected in the values of sets of low-energy coupling constants.
The piece ${\cal L}^{sources}_{QCD}$ will be treated perturbatively.

For the moment, 
let us  examine the symmetry properties of ${\cal L}^0_{QCD}$. We recall
the basic properties of the Dirac gamma matrices
\be
\{\gamma_\mu, \gamma_\nu\}=2 g_{\mu\nu},\quad \{\gamma_\mu,\gamma_5\}=0
\en 
A particular representation for these matrices is
\be\lbl{gamma}
\gamma_0=\left(\begin{array}{cc}
0 & -\mathbf{1}\\
-\mathbf{1}& 0 \\
\end{array}\right),\ 
\gamma_i=\left(\begin{array}{cc}
0 & \sigma_i\\
-\sigma_i& 0 \\
\end{array}\right),\ 
\gamma_5=\left(\begin{array}{cc}
\mathbf{1} & 0\\
0 & -\mathbf{1} \\
\end{array}\right)\ .
\en
The entries there are 2x2 matrices and $\sigma_i$ are the three Pauli 
matrices. Chiral symmetry is a group of transformations 
which acts on the light quark fields.
The heavy quark fields as well as the gluon fields are unaffected by these
transformations. In order to see how these transformations operate let us
form a $N_F^0$--components 
vector from the  $N_F^0$ light quarks, with $N_F^0=3$, for instance
\be\lbl{psiflav}
\psi=\left(\begin{array}{c}
u\\
d\\
s\\   \end{array}
\right)\ .
\en   
Next, let us introduce two projectors
\be
P_L= {1-\gamma^5\over2},\quad
P_R= {1+\gamma^5\over2}\ .
\en
It is easily verified that the projectors satisfy the following properties
\bea\lbl{projop}
&& P_L+P_R=1,\quad P_L^2=P_L,\quad P_R^2=P_R\nonumber\\
&& P_LP_R=P_RP_L=0, \quad \gamma_\mu P_L=P_R \gamma_\mu,\quad
\gamma_\mu P_R=P_L \gamma_\mu
\ena
With the help of these we define projected spinor fields
\be
\psi_L= P_L\psi,\quad \psi_R= P_R\psi
\en
and, using that $\bar\psi=\psi^\dagger\gamma^0$
\be
\bar\psi_L= \bar\psi P_R,\quad \bar\psi_R= \bar\psi P_L\ .
\en
The projected spinors are eigenstates of ``chirality''
\be
\gamma^5\psi_L =-\psi_L,\quad \gamma^5\psi_R =\psi_R\ .
\en
Using the representation \rf{gamma} 
above for the gamma matrices these projected
spinors can be associated with two-component spinors $\tilde\psi_L$
and $\tilde\psi_R$ (called Weyl spinors)
\be
\psi_L=\left(\begin{array}{c}
0\\
\tilde\psi_L\\ \end{array}\right)
\quad
\psi_R=\left(\begin{array}{c}
\tilde\psi_R\\
0           \\ \end{array}\right)
\en
In the absence of interactions the projected spinors  are also eigenstates of 
helicity. Starting from  the free Dirac equation, 
and considering positive-energy plane-waves $\tilde\psi_L(x)= \exp(-ipx)
\tilde\psi_L(p)$, 
$\tilde\psi_R(x)= \exp(-ipx) \tilde\psi_R(p)$, 
indeed, one easily finds
that $\tilde\psi_L(p)$ and $\tilde\psi_R(p)$ 
correspond to  states with a left-handed and a right-handed 
helicity respectively,
\be
\vec\sigma \cdot \widehat{ p}\, \tilde\psi_L(p)= - \tilde\psi_L(p)\ ,\quad
\vec\sigma \cdot \widehat{ p}\, \tilde\psi_R(p)=   \tilde\psi_R(p)\ .
\en
The key feature of ${\cal L}^0_{QCD}$ is that, using eqs.\rf{projop}, 
we can rewrite the massless  
quark part as two independent pieces in terms of $\psi_L$ and $\psi_R$,
\be\lbl{LLRR}
{\cal L}_{QCD}^{chir}\equiv 
\bar\psi i\gamma^\mu D_\mu\psi 
= \bar\psi (P_L+P_R) i\gamma^\mu D_\mu (P_L+P_R)\psi 
={\cal L}_{QCD}^L + {\cal L}_{QCD}^R
\en
with
\be
{\cal L}_{QCD}^L=\bar\psi_L i\gamma^\mu D_\mu\psi_L,\quad 
{\cal L}_{QCD}^R=\bar\psi_R i\gamma^\mu D_\mu\psi_R\ .
\en  
It is important to note that the decomposition \rf{LLRR} holds in QCD because
the gluons are vector particles (and not scalars or tensors).  Let $g_L$ be
a $N_F^0\times N_F^0$ matrix and let us perform the transformation 
\be\lbl{transL}
\psi_L\to \,g_L \psi_L,\ \bar\psi_L\to \bar\psi_L\,g_L^\dagger
\en
(we recall that according to eq.\rf{psiflav} $\psi_L$ is a $N_F^0$ components
vector)
while $\psi_R$ is unaffected. We see that the Lagrangian \rf{LLRR} is left
invariant by this transformation provided that the matrices $g_L$ satisfy
\be
g_L^\dagger\, g_L= I\ . 
\en
In other terms $g_L$ must be a unitary matrix. The Lagrangian \rf{LLRR} is
thus found to be invariant under transformations by a group of unitary matrices
which we may call $U_L(3)$ acting on $\psi_L$. 
Obviously,  the Lagrangian is also invariant under a group
of unitary matrices acting on $\psi_R$. The complete invariance group
of ${\cal L}^0_{QCD}$ is therefore
\be
U_L(N_F^0)\times U_R(N_F^0)\ .
\en
We can parametrize an arbitrary element of this group in the following way
\be\lbl{paramg}
(g_L,g_R)=(\e^{i\alpha} \e^{i\beta} \e^{i\alpha^l T^l} \e^{i\beta^l T^l},
\e^{i\alpha} \e^{-i\beta} \e^{i\alpha^l T^l} \e^{-i\beta^l T^l} )
\en
in terms of $2(N_F^0)^2$ real parameters. 
This expression  shows that one can factor out two $U(1)$ groups, 
such that one can write
\be
U_L(N_F^0)\times U_R(N_F^0)= U_V(1)\times U_A(1)\times SU_L(N_F^0)
\times SU_R(N_F^0)\ .
\en
One can also verify that the ensemble of matrices which have the form
\be
(g,g)=(\e^{i\alpha^l T^l} ,\e^{i\alpha^l T^l} ) 
\en
form a subgroup which we 
will label as $SU_V(N_F^0)$. The set of matrices of the form 
\be
(h,h^\dagger)=(\e^{i\beta^l T^l} ,\e^{-i\beta^l T^l} ) 
\en
are called axial transformations, this set does {\sl not} form a group. 
It is useful to note how the spinor $\psi=\psi_L+\psi_R$ transforms
under a vector and an axial transformation. Starting from eq.\rf{transL}
and the similar one for $\psi_R$ it is easy to deduce
\be 
V:\psi\to \e^{i\alpha^l T^l}\psi \quad\ 
A:\psi\to \e^{i\beta^l T^l \gamma^5}\psi \ .
\en
\subsection{Consequences of the symmetries}
According to Noether's theorem, to each parameter of a continuous symmetry
corresponds a current $J_\mu(x)$ which is conserved, i.e. satisfies
$\partial^\mu J_\mu(x)=0$ (see e.g. \cite{weinbook}). 
%citer une reference ?
The method for obtaining  the explicit form of the current is standard. In the
present case, we consider $\psi(x)$ fields which are solutions
of the classical equations of motion, such that the action is invariant under 
arbitrary infinitesimal variations $\delta\psi(x)$ to
first order. Then we generate variations associated with chiral 
transformations with infinitesimal values of the parameters $\alpha$, $\beta$, 
$\alpha_l$, $\beta_l$, taken to be $x$ dependent. For instance, in order to
obtain the axial current we perform the variation
\be
\delta\psi(x)= i\beta^l(x) T^l \gamma^5 \psi(x),\quad 
\en 
for arbitrary infinitesimal $\beta^l(x)$ and use the fact that 
$\delta S^0_{QCD}=0$. 
In this way one easily 
obtains the form of the vector and axial-vector currents,
\be
V_\mu^0=\bar\psi\gamma_\mu\psi,\quad
V_\mu^a=\bar\psi\gamma_\mu{\lambda^a\over2}\psi,\quad
A_\mu^a=\bar\psi\gamma_\mu \gamma_5 {\lambda^a\over2}\psi,\quad\ .
\en
The singlet axial-vector current $A_\mu^0$  is also conserved at the 
classical level, but the conservation law $\partial^\mu A_\mu^0=0$ turns out
to be modified in $QCD$ by quantum mechanical effects 
(this is the famous ABJ anomaly\cite{ABJ} ). This  was analyzed in detail  
by 't Hooft ref.\cite{thooft1} and the result is that the 
$U_A(1)$ group is not a symmetry
of ${\cal L}^0_{QCD}$ at the quantum level.

In addition to this continuous symmetry group, the
$QCD$ action is invariant under the discrete symmetries of parity, charge
conjugation and time reversal invariance\footnote{In principle, a so-called
$\theta$-term which has the form ${\cal L}=\theta 
\epsilon^{\mu\nu\rho\sigma} G^a_{\mu\nu} G^a_{\rho\sigma} $ is allowed. 
This term is not invariant under $CP$. }.
%citer coleman ? 
Starting from the 
representation in terms of Weyl spinors, 
\be
\psi(t,\vec{x})=\left(\begin{array}{l}
\tilde\psi_R (t,\vec{x})\\
\tilde\psi_L (t,\vec{x})\end{array}\right)
\en
let us recall how
$P$ and $C$, for instance, act (see e.g. \cite{weinbook}),
\be
P:\psi\to \left(\begin{array}{l}
\tilde\psi_L (t,-\vec{x})\\
\tilde\psi_R (t,-\vec{x})\end{array}\right),\quad
C:\psi\to \left(\begin{array}{l}
\sigma_2\tilde\psi_L^* (t,\vec{x})\\
-\sigma_2\tilde\psi_R^* (t,\vec{x})\end{array}\right)
\en
The action of both $C$ and $P$ acting on the chiral part of the QCD 
Lagrangian (see \rf{LLRR} ),
${\cal L}^{chir}_{QCD}$ is to 
{\sl interchange} the parts ${\cal L}^L_{QCD}$ and ${\cal L}^R_{QCD}$. 

At this point we may start to ask ourselves about the consequences of this
symmetry concerning the spectrum of bound states of QCD. 
%better discussion
The important issue will be to know whether the ground state
of QCD (i.e. the vacuum) is also invariant under the full chiral group or not.
Let us first assume that the vacuum is fully invariant: the expectation then
(still ignoring the small effect of light quark masses) is that the spectrum
will consist of degenerate double $SU(3)$ multiplets of opposite parity. Let
us show this, for example, in the case of the vector and the axial vector
mesons. Let us consider the correlator
\be
\Pi_{\mu\nu}^{+-}=<0\vert T V^+_\mu(x) V^-_\nu(0) \vert 0>,\quad
V^\pm _\mu(x)= \bar\psi(x) {\lambda_1\pm i\lambda_2\over 2}
\gamma^\mu \psi(x)\ .
\en
Firstly, using invariance of the vacuum under $SU_V(3)$ transformations
\be
{\rm e}^{i\alpha^l Q^l_V}\vert 0 >= \vert 0 >
\en
one can derive that the correlators are flavour symmetric, i.e. 
for any two flavour indices $i,j$ the correlator must be of the form
\be
\Pi_{\mu\nu}^{ij}= {1\over 2}\delta^{ij}  \Pi_{\mu\nu}^{+-}\ .
\en
Next, let us also assume
invariance of the vacuum under the axial transformation
\be\lbl{axinvar}
{\rm e}^{i\beta Q^3_A}\vert 0 >= \vert 0 >\ .
\en
This implies
\be\lbl{transcor}
<0\vert T V^+_\mu(x) V^-_\nu(0) \vert 0>= 
<0\vert {\rm e}^{-i\beta Q^3_A}T V^+_\mu(x) V^-_\nu(0) 
{\rm e}^{i\beta Q^3_A}\vert 0>= 
<0\vert T V^{'+}_\mu(x) V^{'-}_\nu(0) \vert 0>
\en
where
\be
 V^{'+}_\mu(x)=\bar\psi'(x) {\lambda_1\pm i\lambda_2\over 2}
\gamma^\mu \psi'(x),
\quad \psi'(x)={\rm e}^{i\beta {\lambda^3\over2}\gamma^5}\psi(x)\ .
\en
Expanding up to quadratic order in the parameter $\beta$ we get
\bea
&& V^{'+}_\mu(x)=  V^{+}_\mu(x)+i\beta A^{+}_\mu(x) -{1\over2}\beta^2 V^{+}_\mu(x)
\nonumber\\
&& V^{'-}_\mu(x)=  V^{-}_\mu(x)-i\beta A^{-}_\mu(x) -{1\over2}\beta^2 V^{-}_\mu(x)\ .
\ena
Replacing in eq.\rf{transcor} and collecting the terms quadratic in $\beta$
we obtain that the correlator of vector currents is identical to the
correlator of axial-vector currents
\be\lbl{corrdeg}
<0\vert T V^+_\mu(x) V^-_\nu(0) \vert 0>=
<0\vert T A^+_\mu(x) A^-_\nu(0) \vert 0>\ .
\en
Eq.\rf{corrdeg} implies that the vector and axial-vector mesons should form 
mass degenerate multiplets. One could repeat the same argument using scalar and
pseudo-scalar currents baryon currents etc...
However, this is
not what is observed experimentally: while such multiplets exist, the 
axial-vectors are much heavier than the vectors. 
The reason for this behaviour is
that while the QCD Lagrangian is invariant under the chiral symmetry group
the ground state of the theory is not invariant under the full group
(one says that there is {\sl spontaneous breakdown} of this part 
of the symmetry group). The 
possibility of such a behaviour in field theory was pointed out by Nambu and
by Goldstone\cite{NG}.  An important implication which  they noted is 
the related existence of {\sl massless} particles, now referred to as 
the Nambu-Goldstone bosons.  
We will illustrate this property, which is crucial
for the low energy properties of QCD and the chiral expansion method, in the
next section. Before that, let me simply state 
a set of nice  results
which have been proved in the case of QCD. \\
\indent a) The subgroup $SU_V(N_F^0)$ of
the chiral group cannot be spontaneously broken\cite{vafa_witten}.\\ 
\indent b) The lightest particles in the QCD spectrum are  pseudoscalar 
mesons\cite{weingarten}\\
\indent c) Assuming that QCD with $N^0_F$ massless quarks 
confines, then, in the case $N^0_F\ge 3$, one can show that the  
QCD spectrum {\sl must} contain massless boson states
in order to properly satisfy the so called anomaly matching 
conditions\cite{thooft}.  
These bosons, according to the result b) must be pseudoscalar bosons. 
This last result implies that symmetry under the axial transformations
must undergo spontaneous breakdown in QCD. It is however not excluded
that a discrete axial subgroup could remain unbroken.

\section{Illustration of spontaneous symmetry breaking}
\subsection{A spin model}
We first illustrate with a simple lattice model that 
in a system with an infinite number of variables, the possibility  
that spontaneous breakdown of a symmetry  occurs is rather natural.
Let us consider a system of spins, represented as 
3-vectors of unit length, 
located  on the sites of an infinite dimensional lattice. The
Hamiltonian of this system is taken to be
\be\lbl{spinham}
H=J \sum  \vec S_i . \vec S_j 
\en
where the sum extends, say, over nearest neighbours.
The Hamiltonian is invariant  under the  transformations 
$\vec S_i \to g \vec S_i$ which leave the scalar product invariant, i.e.
$g$ belongs to 
the group of 3-dimensional rotations $O(3)$.   
Concerning the ground state of the
system, we have two possibilities
\begin{itemize} 
\item[a)\ ] the system is disordered 
\item[b)\ ] the system is ordered. 
\end{itemize}
As an example of the latter possibility we may have all
the spins aligned along a given direction which would correspond to a 
ferromagnetic system. This is illustrated in fig.1 below.
%------ 
\begin{figure}[ht]
\begin{center}
\includegraphics[width=6cm]{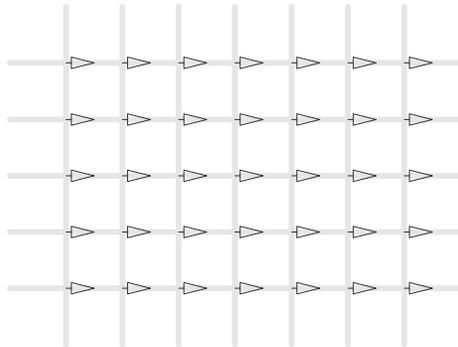}
\caption{\sl Ordered ground state corresponding to a ferromagnetic system.}
\end{center}
\end{figure}
%-----
In this case, the ground  state is not invariant under the symmetry group
$O(3)$ it is invariant under the subgroup of rotations perpendicular
to the $x-$axis, that is, an $O(2)$ group. Alternatively, the ground state of
the system may be disordered, i.e. the directions of the spins are randomly
distributed. This is illustrated in fig.2. 
%------ 
\begin{figure}[ht]
\begin{center}
\includegraphics[width=6cm]{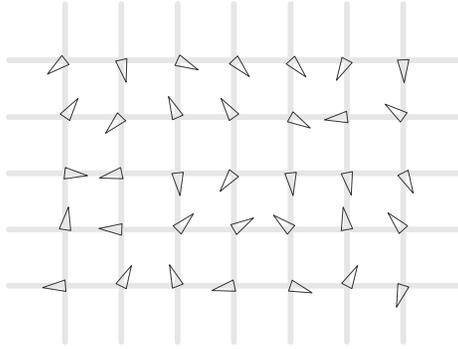}
\caption{\sl Disordered ground state.}
\end{center}
\end{figure}
%----- 
In this case, the aspect of the system will appear to be the same for an
observer in any reference frame. In particular, the ground state is invariant
under the same symmetry group $O(3)$ as the Hamiltonian \rf{spinham}.
We may characterize the nature of the ground state by
an order parameter. For instance, we may choose the average value of the
spin along the $x-$axis as an order parameter
\be
 M_1=<  S^x >={1\over N}\sum_i {S^x_i}
\en
Obviously, $M_1$ vanishes  if the state is disordered, as in fig.2 and it
is non-vanishing in the case of the ferromagnetic state. 
More complicated forms of order are also possible. 
For instance, fig.3 illustrates
the case of an anti-ferromgnetic ground state.
%------ 
\begin{figure}[ht]
\begin{center}
\includegraphics[width=6cm]{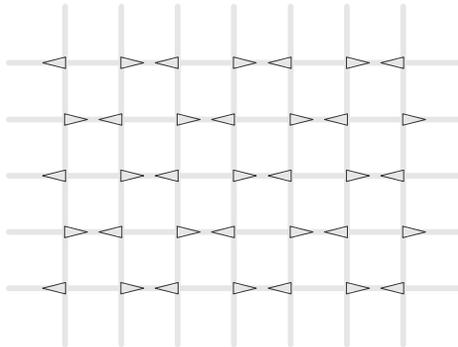}
\caption{\sl Ordered ground state corresponding to a anti-ferromagnetic system.}
\end{center}
\end{figure}
%-----
In this case, the state is invariant under the group $O(2)$ of rotations
perpendicular to the $x-$axis and also under the change of $x$ into $-x$
so the full symmetry group is $O(2)\times Z_2$. The order parameter $M_1$
vanishes for this state but we can imagine a more complicated  order parameter
\be
M_2 = {1\over N}\sum S^x_i S^x_{i+1}
\en
which does not, while $M_2$ would still vanish in the case of a disordered
ground state. Analogous situations in the case of QCD were discussed in 
refs.\cite{stern0} and \cite{kks}.

\subsection{The linear sigma model}
Now we consider a continuous model, the classic linear sigma model\cite{lsm}
and we will investigate in more details the dynamics of the fluctuations
around the ground state. The dynamical degrees of freedom consist
of four real-valued scalar fields $\phi_a(x)$ which can be collected
into a four-component vector $\tilde\phi(x)$ and the dynamics is defined
from the following Lagrangian,
\be
{\cal L}_{lsm}={1\over 2} \partial_\mu\tilde\phi(x)\cdot 
\partial^\mu\tilde\phi(x) + {\mu\over2}  \tilde\phi(x)\cdot \tilde\phi(x)
-{\lambda\over4} (\tilde\phi(x)\cdot \tilde\phi(x))^2
\en
One must have $\lambda>0$ otherwise the Hamiltonian would not be bounded
from below and $\mu$ can have either sign. The Lagrangian is invariant under
transformations of the four-vector $\tilde\phi$
\be\lbl{o4}
\tilde\phi\to G \tilde\phi
\en 
where $G$ belongs to the orthogonal group $O(4)$.  One can 
reformulate the model such that the global 
symmetry resembles the one in QCD. For this purpose, let us map the four 
vector to a $2\times2$ matrix $\Sigma$
\be
\Sigma=\phi_0 +i\vec\tau \cdot \vec\phi\ .
\en
In terms of $\Sigma$ the Lagrangian can be written as follows,
\be\lbl{lsm}
{\cal L}_{lsm}={1\over 4}\tr(\partial_\mu\Sigma\partial^\mu\Sigma^\dagger)
+{\mu\over4}        \tr(\Sigma\Sigma^\dagger) 
-{\lambda\over16}(\tr(\Sigma\Sigma^\dagger) )^2\ .
\en
In this form, the invariance group appears to be  $SU(2)\times SU(2)$ with
\footnote{A one-to-one mapping exists between $O(4)$ and
$SU(2)\times SU(2)/Z_2$ since the kernel of the transformations \rf{o4}
reduces to the identity while the kernel of the transformations \rf{su22}
is a group $Z_2$ consisting of the elements $(1,1)$ and $(-1,-1)$.}
\be\lbl{su22}
\Sigma\to g \Sigma h^\dagger 
\en
which resembles the case of $QCD$ with $N^0_F=2$. 

Let us now consider the ground state $\psi_0$ of this system. By analogy
with the spin system considered above we may choose the expectation values
of the components of $\tilde\phi$ as order parameters. By a suitable 
redefinition we can always arrange that the expectation value of the $\phi_0$
component
\be
<\Psi_0\vert \phi_0 \vert \Psi_0>\equiv \bar\phi_0
\en
may or may not be vanishing while the other expectation values 
$\bar\phi_i, i=1,2,3$ are always
vanishing.
In a semi-classical approximation  $\bar\phi_0$ is given
simply by solving the classical equations of motion for $\phi_0$
(this can be seen easily using the path integral formalism).
Because of translational invariance of the vacuum  $\bar\phi_0$ is independent
of $x$ and the equations of motion simply read
\be
\bar\phi_0 (\mu -\lambda \bar\phi_0^2)=0\ .
\en
If $\mu<0$ the only solution is $\bar\phi_0=0$ which corresponds to a 
disordered vacuum, invariant under the full symmetry group. If $\mu>0$
there are solutions with $\bar\phi_0\ne0$, 
\be
\bar\phi_0=\pm \sqrt{\mu\over\lambda} \ .
\en
Fig. 4 shows that these are
the stable solutions. In this case, the subgroup
$SU_V(2)$ of transformations with elements of the form 
$(g,g)$ leave the vacuum invariant
\be
g \bar\phi_0 g^\dagger =\bar\phi_0
\en
while the set of axial transformations $(h,h^\dagger)$ do not
\be
h\bar\phi_0 h \ne\bar\phi_0\ .
\en 
%-------------------------------------------------------------------
\begin{figure}[hbt]
\begin{center}
\psfrag{energy}{$< H>$ }
\psfrag{phi0bar}{$\bar\phi_0$}
\includegraphics[width=10cm]{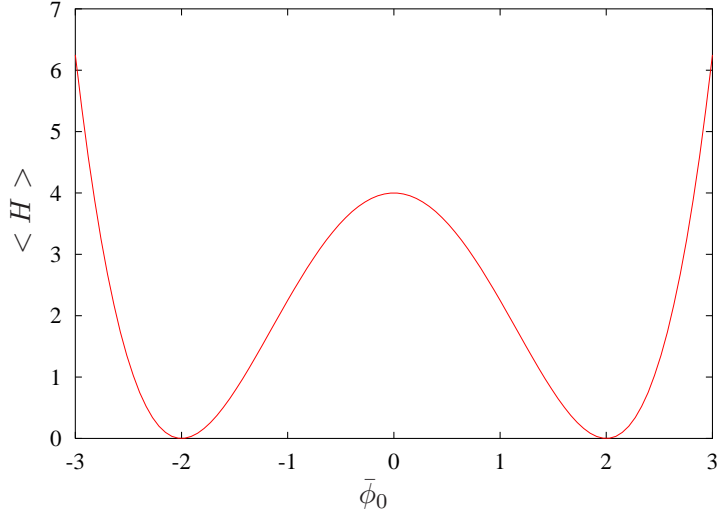}
\caption{\sl Value of the Hamiltonian as a function of $\bar\phi_0$. }
\end{center}
\end{figure}
%-------------------------------------------------------------------
Correspondingly, we expect the appearance
of three massless states: this is Goldstone theorem. The theorem holds
in the absence of gauge fields coupling to these
states and can be proved quite generally \cite{goldstone_theor}. 
We will verify this in the semi-classical approximation: for this purpose, we
consider small fluctuations around the classical solution. One way to perform
this is to write the $\Sigma$ field as,
\be\lbl{paraml}
\Sigma=\bar\phi_0 +\phi_0 +i\vec\tau . \vec\phi\ .
\en
We will choose, however, the following  alternative parametrization 
for the fluctuations
\be\lbl{paramnl}
\Sigma= (\bar\phi_0 +\sigma) U,\quad U=\exp\left(
{i\vec\tau\vec\pi\over\bar\phi_0} \right)\ .
\en
The parametrization \rf{paraml} is the most convenient one if 
we want to perform an ordinary perturbative expansion (i.e. 
an expansion in powers of
the coupling constant $\lambda$). The alternative parametrization will
allow us to perform a radically 
different kind of expansion: the chiral expansion. 
Replacing in the Lagrangian\rf{lsm} one obtains,
\bea\lbl{lsm1}
&& {\cal L}_{lsm}={1\over4}(\bar\phi_0 +\sigma)^2 \trace{\partial_\mu U
\partial^\mu U^\dagger}\\
&&\phantom{{\cal L}_{lsm}=}
+{1\over2}\partial_\mu \sigma \partial^\mu \sigma +
+{\mu\over2} ( \bar\phi_0 +\sigma)^2 -
{\lambda\over4} ( \bar\phi_0 +\sigma)^4\ .
\ena
One observes that no mass term appears in this Lagrangian for the three
pions $\pi_a$, 
this is in accordance Goldstone theorem. In addition,
one has a $\sigma$ meson, which is massive,
\be
m_\sigma^2= 2\mu\ .
\en
One notices from \rf{lsm1} that the strength of the interaction among pions
vanishes for vanishing momenta, such that for small pion momenta we have
a weakly coupled system.
The main idea of the chiral expansion is that if we confine ourselves to a
range of energies (or momenta) for the pions which is much smaller than 
$m_\sigma$ we can ``integrate out'' the sigma field. In this way we 
can describe the dynamics of pions as an expansion in powers of the momenta. 
For this purpose, let us introduce a {\sl chiral counting rule}: we count the
term $\trace{\partial_\mu U \partial^\mu U^\dagger}$ as $O(p^2)$ since it
involves two derivatives of the pion field. How do we count the $\sigma$
field ? We expect it to be a small fluctuation so let us count it as $O(p^2)$
also. These rules automatically give rise to an expansion. To begin with, 
we have a single term of order $O(p^2)$,
\be
{\cal L}^{(2)}= {1\over4}(\bar\phi_0 )^2 \trace{\partial_\mu U
\partial^\mu U^\dagger}\ .
\en 
At $O(p^4)$, at the classical level we must keep terms linear and quadratic
in the sigma field,
\be
{\cal L}^{(4)}= {1\over2} \bar\phi_0\sigma \trace{\partial_\mu U
\partial^\mu U^\dagger} -{1\over2} m_\sigma^2 \sigma^2\ .
\en
It is easy here to integrate over the $\sigma$ field and we obtain
\be
{\cal L}^{(4)}= {(\bar\phi_0 )^2\over 8m_\sigma^2 }\trace{\partial_\mu U
\partial^\mu U^\dagger}^2\ . 
\en
From this, we see that we are generating an expansion in powers of 
$p^2/ m_\sigma^2$. We have performed here a simplified expansion at the 
classical level, we will mention below 
how quantum effects (i.e. loops) are to be taken into account. The possibility
of performing a chiral expansion is very general and only requires that 
a continuous symmetry is spontaneously broken. In the next section,
we move to the more interesting case of QCD.

\section{Low-energy expansion in $QCD$}
\subsection{Order parameters}
We now return to $QCD$, let us identify some operators which can be
considered as order parameters. By analogy with the case of the ferromagnet
such operators must be non-invariant under a chiral transformation. The
simplest such operator is the so-called quark condensate,
\be
{\cal Q}_u= \bar u_L u_R +\bar u_R u_L =\bar uu\ .
\en
As a result of the spontaneous breaking of axial symmetry, the vacuum
expectation value 
\be
<\Psi_0\vert {\cal Q}_u\vert \Psi_0>\ne 0 
\en
is expected to be non-vanishing and by flavour symmetry, if $N_F^0=3$, 
one has 
 $<\Psi_0\vert\bar uu\vert\Psi_0>$ = 
 $<\Psi_0\vert\bar dd\vert\Psi_0>$ =
 $<\Psi_0\vert\bar ss\vert\Psi_0>$. 
By analogy with the case of the anti-ferromagnet discussed in sec. 3
the quark condensate will vanish if  a discrete subgroup
of the axial transformations happened to remain unbroken,
for example the subgoup $Z_4$ with elements,
\be\lbl{Z4}
(g_k,g^\dagger_k),\quad g_k= \exp({ik\pi\over 2}\lambda_3 ),k=0,1,2,3.
\en
Such a possibility was discussed in ref.\cite{kks}. The authors 
argued that  it was ruled out in the case of QCD.  
Ruling out such a possibility based on experiment is also interesting. 
This is rather difficult
and has been achieved only recently by comparing 
calculations of $\pi\pi$ scattering in ChPT at NNLO 
with accurate experimental data\cite{nozeroBexp}. 
We can construct many operators which are
order parameters. For instance, 
\be
{\cal Q}^{(4)}=\bar u_L \gamma_\mu d_L  \bar d_R \gamma^\mu u_R
\en
is an  order parameter which is also invariant under the transformations
\rf{Z4} contrary to the operator $\bar uu$.
 
\subsection{Source term}
The  part of the $QCD$ Lagrangian which was labeled 
${\cal L}_{QCD}^{sources}$ 
has so far been ignored. Let us first write this term in a generic way,
in terms of a set of sources $v_\mu$, $a_\mu$, $s$, $p$ which are 
$N^0_F\times N^0_F$ matrices in flavour space
\be
{\cal L}_{QCD}^{sources}= \bar\psi \gamma_\mu ( v_\mu + \gamma_5 a_\mu )
\psi - \bar \psi ( s +i\gamma_5 p ) \psi
\en
The vector and axial-vector sources correspond to physical fields from
the electroweak sector treated at the classical level,
%Formula corrected ( v+a , v-a for the Z were inverted) 
\bea
&&(v_\mu + a_\mu)_{phys} = -e Q F_\mu  -{e\over\sin\theta_W\cos\theta_W}
(-Q \sin^2\theta_W  )\, Z_\mu 
\nonumber\\
&&(v_\mu - a_\mu)_{phys} = -e Q F_\mu -{e\over\sin\theta_W\cos\theta_W} 
(-Q \sin^2\theta_W + Q -{1\over6}I)\,Z_\mu 
\nonumber\\
&& \phantom{ v_\mu - a_\mu)_{phys} = -e Q F_\mu +}
-{e\over\sqrt2 \sin\theta_W}
\left(\begin{array}{ccc}
0              & V_{ud} W^+_\mu & V_{us} W^+_\mu \\
V_{ud} W^-_\mu & 0              &0 \\
V_{us} W^-_\mu & 0              &0 \\ \end{array}\right)
\ena
where $Q=diag(2/3,-1/3,-1/3)$ 
is the quark charge matrix. The scalar sources correspond to the
Higgs field but the only observable effect at low energy 
is via the quark masses
\be
(s+i p)_{phys}= \left( \begin{array}{ccc}
m_u & 0   & 0 \\
0   & m_d & 0 \\
0   & 0   & m_s \\ \end{array}\right)\ .
\en

The source term of the QCD Lagrangian is not invariant under the chiral
group. Indeed, under an infinitesimal vector transformation with parameter
$\alpha\equiv \alpha^l T^l$ (see \rf{paramg}) one finds the following variation
\be
\delta_\alpha {\cal L}_{QCD}^{sources}= 
i\bar\psi\left( \gamma^\mu [v_\mu,\alpha] + 
       \gamma^\mu \gamma_5 [a_\mu,\alpha]  
                         - [s,\alpha] 
                -i\gamma_5 [p,\alpha]\right)\psi \ .
\en
Under an infinitesimal axial transformation with parameter $\beta\equiv 
\beta^l T^l$ one finds the following transformation
\be
\delta_\beta {\cal L}_{QCD}^{sources}= 
-i\bar\psi\left( \gamma^\mu \gamma_5 [v_\mu,\beta]+
                          \gamma^\mu [a_\mu,\beta]
                          +\gamma_5  \{s,\beta\} 
                          +i        \{p,\beta\} \right)\psi \ .
\en
An important physical consequence of the non-invariance of the scalar
source term is that the set of
Nambu-Goldstone bosons acquire masses and become pseudo-Goldstone bosons.
In order to properly incorporate the transformation  behaviour 
of the sources into the effective Lagrangian
one uses the following trick (spurion method): assuming that the sources are
formally allowed to transform under the chiral group what is the transformation
law that leaves the full Lagrangian, including the source term invariant ?
It is easy to see that the sources must transform as follows,
\bea
&& v_\mu-a_\mu \to g_L (v_\mu-a_\mu ) g^\dagger_L\nonumber\\
&& v_\mu+a_\mu \to g_R (v_\mu+a_\mu ) g^\dagger_R\nonumber\\
&& s+ip \to g_R (s+ip ) g^\dagger_L\nonumber\\
&& s-ip \to g_L (s-ip ) g^\dagger_R\
\ena
In fact, once we allow the sources to transform we have an even more
general invariance property.
We find that we can make the full Lagrangian invariant under
{\sl local} chiral transformations, 
provided that we assume that the vector and 
axial-vector sources formally transform as follows,
\bea
&& v_\mu(x)-a_\mu(x) \to g_L(x) (v_\mu(x)-a_\mu(x) ) g^\dagger_L(x)
+ i g_L(x) \partial_\mu g^\dagger_L(x) \nonumber\\
&& v_\mu(x)+a_\mu(x) \to g_R(x) (v_\mu(x)+a_\mu(x) ) g^\dagger_R(x)
+ i g_R (x)\partial_\mu g^\dagger_R(x) \ .
\ena
Let us  simply mention that this invariance is a property 
of the partition function
of QCD, $ W(v_\mu,a_\mu,s,p )$ which is expressed as a functional integral,
\be
\exp iW(v_\mu,a_\mu,s,p) =
\int d\mu(G,q_f,\bar q_f) \exp i\left( S^0_{QCD} + S^{sources}_{QCD} 
\right) \ .
\en
The invariance under the axial transformations, which holds at the classical
level,  must be modified at the quantum level 
to account for the Adler-Bardeen anomaly 
(see e.g.\cite{gl85} ). Since the QCD Green's functions are obtained 
by taking functional derivatives of  $ W(v_\mu,a_\mu,s,p )$ with respect to
the sources one obtains relations among various Green's functions 
known as chiral Ward identities.  

\subsection{Low-energy effective Lagrangian}
We can now start to construct the effective Lagrangian describing the 
low-energy  dynamics of QCD taking the external sources into account. We must
first provide chiral counting rules for the sources. 
This will allow us to generate a perturbation expansion 
as a function of the momenta and also to expand simultaneously perturbatively 
in terms of the sources.
The obvious choice for
the vector and axial vector sources is to count them as $O(p)$. The choice
for the scalar and pseudoscalar sources is somewhat less obvious. The 
standard choice is is to count them as $O(p^2)$. If the quark condensate were
vanishing or very small, then,  another choice should be made\cite{stern}. 
Instructed by our experience with the linear sigma model we encode the
set of $(N^0_F)^2-1$ pseudo-Goldstone bosons into a unitary matrix U,
\be
U=\exp {i\pi^l \lambda^l \over F_0 }\ .
\en
More explicitly, one has
\be
\pi^l \lambda^l \equiv\left(\begin{array}{lll}
\pi_3 +{1\over\sqrt3}\eta_8 & \sqrt2\pi^+          &\sqrt2 K^+\\
\sqrt2\pi^-           &-\pi_3 +{1\over\sqrt3}\eta_8 &\sqrt2 K^0 \\
\sqrt2 K^-            &\sqrt2\bar K^0        &{-2\over\sqrt3}\eta_8\\
\end{array}\right)
\en
The matrix $U$ transforms as
\be
U\to g_R U g_L^\dagger\ .
\en
We also note the transformation rules under the discrete symmetries of 
parity and charge conjugation
\bea
&&P:\quad U(x)\to U^\dagger(\tilde x)\nonumber\\
&&C:\quad U(x)\to ^t U(x)
\ena
In order to enforce the invariance under local chiral transformation in the
presence of sources we construct a covariant derivative which transforms in
the same way as U,
\be
D_\mu U = \partial_\mu U -i(v_\mu+a_\mu) U +i U (v_\mu- a_\mu)
\en
We can finally write down the chiral Lagrangian of order $p^2$ which
contains two independent terms which satisfy the local chiral invariance 
constraints,
\be\lbl{lag2}
{\cal L }^{(2)}= {F_0^2 \over 4 }\left(
      \trace{ D_\mu U D^\mu U^\dagger}
+2B_0 \trace{ U(s-ip) +U^\dagger(s+ip)} \right)
\en
At leading order in ChPT one computes the QCD Green's 
functions  from the Lagrangian \rf{lag2} at tree level. 
In other terms, we have the following representation for the
QCD partition function $W$,
\be\lbl{Wp2}
W(v_\mu,a_\mu,s,p)={\cal L}^{(2)} (U_{class},v_\mu,a_\mu,s,p ) + O(p^4)
\en
where $U_{class}$ stands for the solution for the $U$ field of the 
classical equations of motion.

\subsection{Interpretation of the low-energy parameters}
The Lagrangian contains two parameters $F_0$ and $B_0$ which values
encode the exact dynamics from the part ${\cal L}^0_{QCD}$ of the
QCD Lagrangian. 
The parameter $B_0$ must depend
on the QCD renormalization scale $\mu_0$ such that the product 
$B_0(\mu_0) m_f(\mu_0) $ 
is invariant. What is its physical interpretation ? To answer this
question we note that the VEV of the operator $\bar u u$ can be expressed
as a derivative of the partition function $W$,
\be
<\Psi_0\vert \bar u u\vert \Psi_0>=- {\delta 
W(v_\mu=0,a_\mu=0,s=m_u,p=0)\over
\delta m_u}\Big\vert_{m_u=0}
\en
We can now compute this same quantity from the low energy effective theory
using eq.\rf{Wp2} and we find
\be\lbl{B0}
<\Psi_0\vert \bar u u\vert \Psi_0>= - B_0 F_0^2\ .
\en
We can also interpret the parameter $F_0$ by computing the matrix element
of the axial current between the vacuum and a one pion state and we find
\be
<\Psi_0\vert \bar\psi(x){\lambda^3\over2}\gamma_\mu\gamma_5
\psi(x)\vert \pi^a(p)>
= i\delta^{3a} p_\mu \e^{-ipx} F_0\ .
\en
Therefore $F_0$ is the so-called pion decay constant which, physically, can
be determined from the decay rate of the charged pions
\be
\Gamma_{\pi^+\to l^+\nu }={G_F^2 \over4\pi} F_0^2 m_l^2 m_\pi V^2_{ud}
\left( 1-{m_l^2\over m_\pi^2}\right) 
\en
We note that $F_0$ is a quantity which is defined in the chiral limit,
it is equal to the physical decay constant only at the leading order
in ChPT. $B_0$ is seen from eq.\rf{B0} to be related
to an operator which is an order parameter. The same property holds for $F_0$
which can be expressed as an integral,
\be
F_0^2 ={1\over6} \int d^4x <\Psi_0 \vert \bar u_L(x) \gamma_\mu d_L(x)\,
\bar d_R(0) \gamma^\mu u_R(0) \vert\Psi_0 >
\en
which involves a correlation function which is an order parameter
of chiral symmetry. This is a rather generic feature of low-energy 
parameters. 
\subsection{Higher orders}
At this point we have defined the chiral expansion at leader order $O(p^2)$.
Let us only briefly  mention here the idea involved in going to the next order
(the details can be found in refs.\cite{weinberg,gl84,gl85}). Firstly, the
Lagrangian ${\cal L}^{(2)}$ must be used not just at the tree level but one
must include the quantum corrections at one loop. Secondly, the chiral 
Lagrangian itself must be enlarged to include a set of terms ${\cal L}^{(4)}$. 
The procedure of renormalization can be shown to be well defined.

\subsection{Quark mass ratios}
Let us now compute, for instance, the masses of the pseudo-scalar mesons.
Expanding the Lagrangian \rf{lag2} to second order in the fields
it is not difficult to obtain,
\bea
& M^2_{\pi^0}= M^2_{\pi^+}&= B_0 (m_u +m_d)\nonumber\\
& M^2_{K^+}             &  =B_0 (m_u +m_s)\nonumber\\
& M^2_{K^0}             &  =B_0 (m_d +m_s)\nonumber\\
& M^2_{\eta}            &  ={B_0\over 3}(m_u +m_d +4 m_s)
\ena
One immediate prediction is the Gell-Mann-Oakes-Renner\cite{gmor} relation
\be
3M^2_\eta= 2(M^2_{K^+}+M^2_{K^0} )- M^2_{\pi^0}
\en
which is rather well satisfied experimentally. Another outcome concerns 
the quark mass ratios. We can first determine the ratio of the strange
over non-strange average mass,
\be
{2m_s\over m_u+m_d}={M^2_{K^+} +  M^2_{K^0} \over M^2_{\pi^0}}-1\simeq 26\ .
\en
We can also determine the isospin breaking mass difference $m_u-m_d$ from
the difference $M^2_{K^+} -M^2_{K^0} $ but we must remove the electromagnetic
contribution to this quantity,
\be
{m_u -m_d \over m_u +m_d}= { M^2_{K^+} -M^2_{K^0} 
-(M^2_{K^+} -M^2_{K^0})^{EM}\over M^2_{\pi^0} }
\en
This electromagnetic piece is itself determined in fact by chiral symmetry
at this order (Dashen's theorem\cite{dt} )
\be
(M^2_{K^+} -M^2_{K^0})^{EM}= M^2_{\pi^+}- M^2_{\pi^0}
\en
which allows one to deduce that $m_u/m_d\simeq 0.55 $. 
Improving on these results using higher order ChPT turns out to be more subtle
than one might expect, for a discussion see ref.\cite{leutrev}. 

\section{Concluding remarks}
At the leading order ChPT is a very predictive theory. The leading order
Lagrangian contains four parameters $F_0,\ B_0 m_u,\  B_0 m_d,\  B_0 m_s $
in terms of which a very large number of quantities can be computed: 
pseudo-scalar masses, pseudo-scalar scattering amplitudes like $\pi\pi\to
\pi\pi$, $K\pi\to K\pi$, the strong decay $\eta\to3\pi$, electromagnetic
form-factors, electromagnetic decays like $\pi^0\to 2\gamma$, weak decays
like $K\to\pi l\nu$, $K\to\pi\pi l\nu$ etc... Unfortunately, leading
order ChPT is not the whole story. In fact, comparing theory and experiment
at this order gives mixed results: sometimes the agreement is very good and 
sometimes there is a factor of two difference. 
In going to higher order one takes into account both quantum effects (loops)
as well as higher-order chiral Lagrangian terms. 
The extension of ChPT 
at NLO was developed in the classic papers of Gasser and 
Leutwyler\cite{gl84},\cite{gl85} and more recently, the formalism as well
as many calculations have been extended to the NNLO level
(see \cite{bijrev}). 
At higher orders many new parameters appear, this limits the predictivity
but one then aims at high precision. One example 
of a successful prediction is the pion-pion S-wave scattering length. 
This quantity can be determined experimentally from the semi-leptonic 
$K$ decay amplitude $K\to\pi\pi l\nu$ and the most recent determination
gives\cite{E865}
\be
a_0^0 = 0.228\pm0.012              \quad ({\rm experiment})
\en
The prediction from the recent NNLO calculations\cite{cola00}  give
\be
a_0^0 = 0.220\pm0.005                \quad ({\rm theory})
\en
which agrees  with experiment and improves considerably over
the leading order prediction $a_0^0=0.16$. 

We have restricted the discussion to the mesonic sector, some results on 
the quark mass ratios were given, other parameters from the electro-weak 
Lagrangian that one may be willing to extract from low energy data are 
the Kobayashi--Maskawa matrix elements $V_{ud}$, $V_{us}$. Chiral symmetry
also influences the interactions of the pseudo--scalars with other particles. 
Much work was devoted to the sectors with one, two or several baryons.
Also, since the chiral expansion parametrizes the way in which physical
observables would vary if we were able to vary the quark masses there are
potential useful applications for performing extrapolations in connection with
lattice QCD simulations.

\end{document}